# Prestige of scholarly book publishers – an investigation into criteria, processes, and practices across countries

Eleonora Dagienė
Centre for Science and Technology Studies (CWTS), Leiden University, the Netherlands
Institute of Communication, Mykolas Romeris University, Vilnius, Lithuania

**Abstract**

Numerous national research assessment policies set the goal of promoting 'excellence' and incentivise scholars to publish their research in the most prestigious journals or with the most prestigious book publishers. We investigate the practicalities of the assessment of book outputs based on the prestige of book publishers (Denmark, Finland, Flanders, Lithuania, Norway). Additionally, we test whether such judgments are transparent and yield consistent results. We show inconsistencies in the levelling of publishers, such as the same publisher being ranked as prestigious and not-so-prestigious in different states or in consequent years within the same country. Likewise, we find that verification of compliance with the mandatory prerequisites is not always possible because of the lack of transparency. Our findings support doubts about whether the assessment of books based on a judgement about their publisher yields acceptable outcomes. Currently used rankings of publishers focus on evaluating the gatekeeping role of publishers but do not assess other essential stages in scholarly book publishing. Our suggestion for future research is to develop approaches to evaluate books by accounting for the value added to every book at every publishing stage, which is vital for the quality of book outputs from research assessment and scholarly communication perspectives.

**Keywords**: book assessment, rankings of publishers, prestigious publishers, research excellence, book publishers, book indicators

## 1. Introduction

For several decades, policymakers in many countries have incentivised researchers to publish their findings through the 'most prestigious channels' to promote 'excellence'. The narratives of 'excellence' and 'quality' in academia (Lamont 2009; Moore et al. 2017), as well as the academic 'prestige culture' (Fyfe et al. 2017) and 'quantified control' (Burrows 2012) for funding allocation affect the perception of research book publishers' 'quality' and 'prestige'. However, no complementary phenomenological investigation has explored further 'what actually goes on in the murky world of academic preferment' (Cronin & La Barre 2004).

Acquisition librarians were the first who asked researchers to determine the 'quality' of publishers (Lewis 2000; Metz & Stemmer 1996) to support the development of library collections. Academics were surveyed to identify 'quality indicators' for books and their publishers (Giménez-Toledo et al. 2013) so as to assist in creating a ranking of publishers.

Apart from this traditional method, many attempts have emerged to quantify the assessment of books, evaluate their impact, and distinguish publishers: using libcitations (White et al. 2018, 2009), book reviews (Zuccala & van Leeuwen 2011; Zuccala & Robinson-García 2019), or a set of digital indicators (Neville & Henry 2014). The newest qualitative and quantitative initiatives are often experimental (Giménez-Toledo et al. 2015), while efforts to be purely quantitative, in some cases, lack clearly stated policies (Basso et al. 2017; Williams & Galleron 2016).

Almost two decades ago, Nederhof et al. (2001) constructed three indices: 'a quality weight', 'an (inter)national visibility weight', and 'a combined index for publishers' assessing publishers in linguistics. Currently, ranking publishing channels and compiling lists of prestigious publishers are common practices for metric-based funding systems. The first thoroughly documented ranking of publication channels (journals, book series, and publishers), called the Norwegian model by its developer Sivertsen (2018), was implemented in 2005. It is a widely discussed and extensively followed approach. In 2008 and 2012, respectively, Denmark and Finland put into practice similar models, including the identification of top-level publishers. Flanders (Belgium) takes a somewhat different approach by differentiating publishers according to their peer review practices (Giménez-Toledo et al. 2016).

In Lithuania, whose research assessment practices have not been investigated yet, policymakers began using the term 'prestigious publisher' a quarter-century ago, but little is known about Lithuanian book assessment practices. For example, some procedures have been described as 'an essentially bureaucratic decision on what is and what is not a book' (Williams et al. 2018). There is an additional unique element in this process in Lithuania compared with other countries. This is the use of a publisher points system to allocate funding: every fourteen pages of a book or a chapter in an edited volume earns points, and therefore funds, for Lithuanian institutions. This seems to be a unique feature of the Lithuanian system. Verleysen and Engels (2018), for instance, discuss weight ratios of publication types, but they do not suggest the possibility of taking into account the number of pages of a publication.

There is one more distinctive Lithuanian feature in the assessment of *book outputs in the sciences*[1]. Only books published by prestigious foreign publishers earn points (and funds)—and nothing is achieved if a publisher is not prestigious. This has elevated publisher prestige to the utmost importance for Lithuanian institutions. Moreover, while formal national regulations define the notion of a prestigious publisher, the decision whether a specific publisher is prestigious or not depends strongly on the opinion of anonymous experts.

Few empirical studies have investigated book assessment based on publisher judgements. Only a handful of papers examine the Norwegian model (see, e.g. Aagaard et al. 2015) or emphasise challenges in verifying the prerequisites for publishers of academic book outputs (Borghart 2013). Several papers flag the unexpected potential consequences of national performance-based funding systems on research practices (Aagaard 2015; Faggiolani & Solimine 2018; Hammarfelt & de Rijcke 2015; Rowlands & Wright 2019). These studies provide a background for more extensive research, particularly regarding the possible implications of assessing books based on their publishers' prestige.

Numerous policies on research assessment, with the goal of promoting 'excellence,' incentivise scholars to publish their best research in the most prestigious outlets; these policies are defined at national, regional, and institutional levels. At the same time, there is still an ongoing debate in research evaluation studies about the endeavour of ranking book publishers per se (Giménez-Toledo et al. 2015). Nevertheless, how such rankings actually distinguish prestigious book publishers has not yet been examined.

---

[1] We use the term 'the sciences' to refer to all fields of research except for the social sciences and humanities.

In this paper, we intend to answer the following research questions:

(1) What methods are employed in Lithuania to identify prestigious publishers in the assessment of book outputs, and how do these methods differ from the methods used in other countries?

(2) To what extent do assessments of book outputs based on the prestige of book publishers yield consistent results, both over time and between countries?

(3) To what extent is it possible to verify whether book publishers meet the formal prerequisites of national assessments?

Using a mixed-methods approach, we will explore the different ways in which the prestige of publishers is determined. This study will contribute to a deeper understanding of the complexity of the assessment of scholarly books. It will also identify the uncertainties of a process in which books are assessed based on their publisher's prestige.

## 2. Research design

This paper presents a case study that employs a combination of qualitative and quantitative methods. A qualitative document analysis was performed to study two related phenomena: *the assessment of scholarly book publishers* and *the methodologies and practices used to determine prestigious publishers*. We used a snowball method for gathering relevant literature starting from Sivertsen (2018). Research papers, edited volumes, regulations, reports, and grey literature were examined to identify rules and practicalities related to the assessment of scholarly books and their publishers. As an example, the documents, regulations, research papers and other information associated with the Norwegian Publication Indicator—as the best-documented indicator followed by other countries—were obtained from its webpage 'About NPI'[2]. Rankings of publishers and the publication points earned by Norwegian institutions were taken from the Norwegian Register for Scientific Journals, Series and Publishers[3]. Lithuanian legal acts containing the methodologies for the formal evaluation of research produced by research and higher education institutions were obtained from an official database: The Register of Legal Acts of the Republic of Lithuania (TAR)[4].

For the bibliometric investigation, we chose the Lithuanian book outputs submitted by institutions to the national metric-based funding systems from 2005 to 2016. We thoroughly examined bibliographic data on registered books assessed by anonymous panel experts. Then, we identified publishers that were awarded both the highest category (prestigious publisher) and at least one other category (not-so-prestigious or non-prestigious). Such discrepancies in judgements about publishers have been discussed within Lithuanian academia at all levels (departments, faculties, and universities) for years. The results are significant for researchers because institutions operate internal incentive schemes reallocating funds received from these annual assessments. However, we only studied issues related to publishers' prestige. We did not analyse the incentive schemes at different institutional levels.

---

[2] The Norwegian Publication Indicator (NPI) https://npi.nsd.no/informasjon accessed 16 April 2020
[3] The Norwegian Register for Scientific Journals, Series and Publishers
https://dbh.nsd.uib.no/publiseringskanaler/Forside accessed 16 April 2020
[4] The Register of Legal Acts of the Republic of Lithuania (TAR) https://www.e-tar.lt/portal/en/index

Bibliographic information on Lithuanian book outputs was derived from databases managed by the Lithuanian Research Council: (1) Dynamics of Lithuanian Research Potential[5] for outcomes published from 2004 to 2008; and (2) Reports on Scientific, Arts and other Relevant Activities of Research and Higher Education Institutions[6] for outputs published from 2009 to 2016. From a bibliographical perspective, the compiled records had various shortcomings. So, we enriched the primary bibliographical data with manual searches of the missing details in multiple catalogues (e.g. the National Bibliographic Database by Martynas Mazvydas, National Library of Lithuania[7], the Lithuanian Academic Electronic Library[8], WorldCat catalogue[9]) and on the web.

ISBNs are mandatory prerequisites for scholarly books in almost all countries, including Lithuania, and registrants of ISBNs can be presumed to be responsible for the content they make publicly available. Thus, in the analysed dataset, we recognised the ISBN registrants as the publishers. We chose the Global Register of Publishers (GRP) as a primary, reliable, and free resource about registrants of ISBNs created by the International ISBN Agency[10]. Thus, alongside the enriched bibliographical information for book outputs gathered from the Lithuanian Research Council, we derived further data about publishers of those books from the GRP.

Many papers on scholarly book evaluation focus on the social sciences and humanities, even though in Norway, Finland, and Denmark (countries that rank publishers to assess book outputs) academic book publishers are not divided into academic fields. In this paper, we distinguish between the sciences, the social sciences, and the humanities only in the section on Lithuania because Lithuanian regulations apply stricter requirements in the sciences than in the social sciences and humanities (see Sub-section 3.5).

Many genres of scholarly books are mentioned in research assessment regulations in different countries. Requirements for publishers usually do not depend on book genres, so we do not distinguish types of books in this paper. Instead, we use the general term 'book' for all kinds of scholarly book outputs and edited volumes. However, in the Lithuanian regulations, different requirements were applied for conference proceedings and their publishers. We, therefore, excluded 50 publications records from the initial dataset of 4135 records (Dagienė et al. 2019), as a thorough investigation of experts' comments revealed that they were conference proceedings published as edited volumes.

The final dataset of Lithuanian book outputs (books and edited volumes and no conference proceedings) analysed in this study reflects institutional submissions of 4085 unique titles having ISBN codes published from 2004 to 2016. The experts positively assessed 3712 (out of 4085) reported book outputs and scored them according to their publisher's prestige. In these cases, the publisher was classified as prestigious or not-so-prestigious. The panels rejected the other 373 titles as inappropriate mostly because the publisher was considered non-prestigious.

---

[5] Lietuvos mokslo potencialo dinamika  http://www.mokslas.mii.lt/mokslas/  accessed 16 April 2020
[6] Mokslo ir studijų institucijų mokslinės, meninės ir su jomis susijusios kitos veiklos ataskaita https://mokslas.lmt.lt/INSTITUCIJOS/ accessed 16 April 2020
[7] National bibliographic database. Martynas Mazvydas National Library of Lithuania https://nbdb.libis.lt/showCustomPage.do?showByIdentificator=complexSearch accessed 16 April 2020
[8] The Lithuanian Academic Electronic Library https://www.lvb.lt/ accessed 16 April 2020
[9]  WorldCat is the world's largest library catalogue. see https://www.worldcat.org accessed 16 April 2020
[10] The Global Register of Publishers https://grp.isbn-international.org/ accessed 16 April 2020

The analysis in this paper focuses on book publishers for which the experts were not consistent, i.e. publishers which some experts classified as prestigious. In contrast, other experts considered them to be not-so-prestigious or even not prestigious at all.

## 3. Defining prestigious book publishers

Numerous research papers confirm that modern research evaluation systems increase the pressure on researchers to publish more and reinforce their 'publish or perish' habits, which significantly changes the publishing patterns of both journal papers and scholarly books (Broz & Stöckelová 2018; Butler 2003; Elton 2000; Good et al. 2015; Moed 2008; Osuna et al. 2011; De Rijcke et al. 2016). Some studies show that scholars adjust their behaviour in response to these research assessments' requirements, especially when the number of publications is explicitly linked to their research funding.

In a qualitative investigation of literature, we found that the regulations surrounding research assessment best reflect the policymakers and scholarly community's perceptions of the 'quality' of book outputs as well as the 'prestige' of publishers producing research outcomes. In recently announced regulations on research assessment, countries approaching qualitative assessment do not emphasise book publishers' prestige. Meanwhile, countries having metric-based funding systems rank publishers and score submitted book outputs according to the levels awarded by experts to publishers.

*In countries having peer review based systems.* In the UK, experts assess the quality of research outputs (and books as well) by reading the actual submitted books, which institutions select as their best outcomes. The UK Research Excellence Framework (REF) clearly states that these are required as physical books (Rosenberg 2015). According to the REF policies, panel reviewers evaluate three distinct elements for each submission: the quality, the impact beyond academia, and the environment that supports research[11]. Also, there is a statement:

> "53. No sub-panel will make use of journal impact factors, rankings or lists, or the perceived standing of the publisher, in assessing the quality of research outputs."
> (Research Excellence Framework 2012)

Nevertheless, several reports with widespread scope commissioned by the UK's Higher Education Funding Council for England (HEFCE) investigate metrics and possible changes in the assessment process. The Metric Tide (Wilsdon et al. 2015) discusses book-based indicators, among other metrics, and Crossick (2015) examines the issues around open access for monographs. The latter relates to policymakers' intention to mandate open access monographs as book outputs in the REF in 2027 (Lockett 2018).

There are more independent reports on the REF2014 results. In one, Tanner (2016) provides a thorough analysis of publishing data on books submitted across the arts and humanities—the experts assessed 8,513 books produced by 1,180 unique publishers. It worth emphasising that 60 per cent of the publishers made only a single submitted book. Tanner's conclusions include:

> "As far as can be ascertained from the available data, attempting to assess books through a purely quantitative method would be nigh on impossible to do fairly or equitably. […] This

---

[11] Research Excellence Framework
https://webarchive.nationalarchives.gov.uk/20180319165633/http://www.ref.ac.uk/about/whatref/ accessed 13 May 2020

study adds further evidence to the sense that bibliometrics remain a very unhelpful means of analysing books for research excellence." (Tanner 2016)

A general independent review on REF2014 results, widely known as the Stern Review (Stern 2016), includes a recommendation supporting the current peer review based assessment and emphasises that if the metrics are provided to inform the evaluation, they should be used transparently.

Similarly to the UK, currently, it looks like the status of publishers is not a decisive factor in France (Williams & Galleron 2016) or Italy (Basso et al. 2017; Faggiolani & Solimine 2018). Nevertheless, an Italian study has been conducted investigating the possibilities of employing quantitative metrics to assess books (Basso et al. 2017). Researchers conclude that classifying publishers is fraught with difficulty (Williams et al. 2018) and suggest surveying researchers, as was done in Spain (see Sub-section 3.1). Nevertheless, based on a thorough systematic review, Giménez-Toledo et al. (2009) conclude that although there is no simple way to determine the 'prestige' of publishers, this is predominant in the research assessment.

*In countries with metric-based assessment funding system*s, the importance of publishers' prestige varies. While the system in some countries does not judge the publishers, in others, it ranks them into levels from a basic entry level to the most prestigious (Giménez-Toledo et al. 2019). In the Czech Republic, the actual publisher of book outputs was of no importance until 2013, when panel peer review evaluation was introduced (Broz & Stöckelová 2018); and the current formal criteria do not mention the importance of publishers (Government of the Czech Republic 2018). In Poland, researchers could self-publish monographs that meet the formal criteria for metric-based assessment funding (Kulczycki 2018) until the List of Scientific Publishers was introduced in 2018 (see Sub-section 3.4).

There are countries with extensive experience in the rankings of publishers. Norway introduced the first and widely documented ranking of publishing channels in 2005 (Sivertsen 2018), Denmark implemented a similar levelling in 2008[12], and Finland followed them with a national system launched in 2012[13]. More details of these rankings are presented in Sub-sections 3.2 and 3.3 below.

Meanwhile, a different publisher assessment system was established in Flanders, the Northern Dutch-speaking region of Belgium. The Flemish regulations do not mention the publisher's prestige; the main criterion is peer review—the procedures expected from the book publishers. (Verleysen & Engels 2013). The national Authoritative Panel, which consists of professors affiliated with Flemish universities with expertise covering the social sciences and humanities, is authorised to evaluate publication channels (journals, publishers, and book series) against the criteria stipulated in the regulations (Verleysen et al. 2014). This panel has found the most challenging aspect of its work is to verify the peer review procedures in book output production. In response to the regulations and doubts, the Flemish Publishers Association invented a label the 'guaranteed peer reviewed content.' So, the Flemish list of publishers consists of two publisher types—those who handle peer review for all their published books and those who manage peer review for individual books or book series.[14] Thus, the Flemish system has no concept of 'prestigious' publishers: Springer, Catholic University of America Press, Oxford University Press, or Berg (to name but a few) are all treated the exact same way, even though some of those are more famous than the others.

---

[12] The BFI is an element of the performance-based model for distribution of the new block grants for research to universities. In: Ministry of Higher Education and Science, Denmark https://ufm.dk/en/research-and-innovation/statistics-and-analyses/bibliometric-research-indicator/bfi-rules-and-regulations accessed 25 April 2020.
[13] Publication Forum https://www.julkaisufoorumi.fi/en/publication-forum accessed 25 April 2020.

## 3.1. Scholarly Publisher Indicators in Spain

Based on a thorough examination of methods to assess monographs through their publishers, Giménez-Toledo and Román-Román (2009) concluded that 'there is no one quality indicator which can be considered determinant and by which the quality of the publisher can be established'.

To find out academics' perceptions of what exactly determines the quality of publishers of monographs, about three thousand Spanish researchers were surveyed (Giménez-Toledo et al. 2013). Considerable variations were revealed within the criteria for the 'prestige' of publishers in different scientific fields. As Giménez-Toledo et al. noticed, some of the leading indicators recognised by researchers (peer review, an ongoing trajectory of publications, publishers' monographs being in libraries, and in international databases) partially coincide with those indicated by the Spanish research evaluation agencies: the National Agency for Quality Assessment and Accreditation and the National Commission for the Evaluation of Research Activity. Besides, the surveyed academics pointed out additional indicators of 'prestige', such as good reviews in the best journals (in Prestige of Publisher), an adequate structure of publications (in Quality of Publications), or publishers maintaining a presence in foreign bookstores (in Dissemination and Distribution System of the Publisher) (Table 3 in Giménez-Toledo et al. 2013).

A Spanish information system on publishers entitled the Scholarly Publishers Indicators was created in 2012, later updated in 2014 and 2018 (Giménez Toledo 2018). Scholarly Publishers Indicators covers only Spanish and international publishers that researchers participating in a survey indicated to be among the top ten in their respective field; thus, publisher prestige in Spain is field-specific. Accordingly, the Scholarly Publishers Indicators allows selecting the most highly valued publishers in sixteen disciplines within the social sciences and humanities. Additionally, the Spanish Scholarly Publishers Indicators includes interactive charts: *Manuscripts Selection Processes* (reported by publishers) and *Scholarly Publishers Indicators Expanded* (showing the presence of each book publisher in five information systems)[15].

Scholarly Publishers Indicators are used quite widely in Spain. According to Giménez-Toledo et al. (2016), Spanish assessment agencies use the indicators only as a reference, and their function is to support the decisions of expert panels. Mañana Rodriguez and Pölönen (2018) specify that the information concerning scholarly publishers in Spain 'is supplemented with further review of the individual titles by expert panels in the context of the applicant's CV' and conclude:

---

[14] The Flemish Academic Bibliographic Database for the Social Sciences and Humanities (VABB-SHW) is a database of academic publications from the social sciences and humanities authored by researchers affiliated to Flemish universities. https://www.ecoom.be/en/data-collections/vabb-shw accessed 25 April 2020.

[15] Scholarly Publishers Indicators Expanded shows the presence of each book publisher in five information systems: Book Citation Index (Clarivate Analytics); Scopus Book Titles (Elsevier); Norwegian list (Norwegian categorization of book publishers, used in various European countries); Scholarly Publishers Indicators / Book publishers' prestige (ÍLIA/CSIC Research Group); Finnish list (Finnish categorization of book publishers) http://ilia.cchs.csic.es/SPI/indexEn.html accessed 13 November 2020

> "It must also be said that a ranking of publishers based on 'quality' does not mean that there is always a direct relationship between a high quality book and a high quality publisher. Expert panels therefore need to have access to each individual publication in order to observe this limitation."

Interestingly, Mañana Rodriguez and Pölönen (2018) present a comparison of ratings and lists in Finland and Spain, revealing publishers with different levels in the two countries; thus, there is no consensus on international and prestigious publishers. The results presented in Section 4 strengthen their findings.

### 3.2. Prestigious publishers in Norway

The Norwegian model, developed for indicator-based funding, incentivises researchers to publish in the most prestigious channels within their research area (Sivertsen 2018). This model implies that prominent researchers designate which journals and book series that have met the entry requirements (level 1) are considered prestigious (level 2) in their particular area of sciences.

Along with research papers, we investigated the regulations published on two separate portals within the Norwegian model. One was taken from the Norwegian Publication Indicator (henceforth referred to as the NPI) ('About the Norwegian Publication Indicator' 2020), and the other from the Norwegian Register for Scientific Journals, Series and Publishers (henceforth referred to as the Norwegian Register). Both have interfaces in English and provide extensive information on processes for publisher ranking.

According to mandatory regulations declared in the Norwegian Register, to be registered at level 1 (which is the basic entry level), book publishers must submit for primary evaluation: (1) their ISBN prefix; (2) documentation of their scientific publishing programme (not the editorial board), (3) external peer review procedures (an explanation in a PDF file is enough), and (4) proof of their international or national authors (names and affiliations from the last two years). Figure 1 shows prerequisites in the Norwegian Registry for the entry level 1 and conditions for level 2.

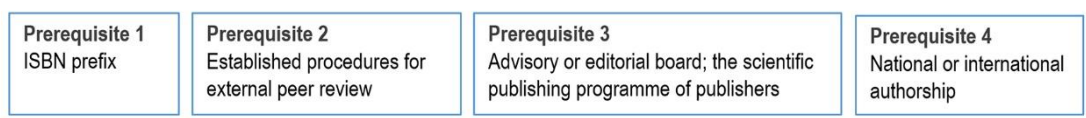

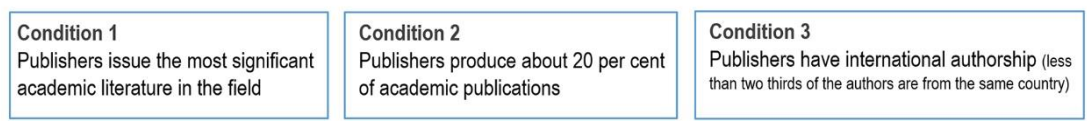

Figure 1. Prerequisites and approval procedures for publication channels of the Norwegian Register

The prestigious level 2 is limited to channels—journals, book series, or book publishers—that issue 'the most outstanding works by researchers from different countries.' While it is unclear how to identify outstanding works, the guidelines state that for calibrating 'prestige'—'level 2 publication channels shall together constitute about one fifth (20 per cent) of the field's total academic publications' (Sivertsen 2018). However, according to the NPI, the Norwegian Register of academic book publishers is not divided into academic fields. The National Board of Scholarly Publishing[16] 'is responsible for the publisher rating levels and updates this annually, based on input from academic fields where book publishing is a central or frequent format for publishing research.' Nevertheless, it seems that level 2 is still limited to leading channels only and has a 20 per cent field-based threshold.

### 3.3. Publisher rankings in Denmark and Finland

Denmark and Finland have implemented the Norwegian model with some adjustments. The main requirements for an entry level listing correspond with those set in Norway (e.g. peer review before publication). Also, as in Norway, local researchers on panels of experts decide which publishers deserve to be designated as prestigious at the highest level.

In Denmark, the Bibliometric Research Indicator[17] (BFI), announced in 2008, has two types of lists: one for journals, books, and conference series, and the other for publishers. These BFI lists were each divided into two levels from 2012 to 2017. Since 2018, these lists have three publication levels: level 1 (ordinary), level 2 (particularly distinguished), and level 3 (prestigious). Level 3 includes no book publishers but only the most prestigious journals, book series and conferences. The allocation of levels depends on the recommendations of researchers who serve on expert panels (*Guidelines for registering research for the Danish Bibliometric Research Indicator* 2019).

The Finnish Publication Forum[18] (henceforth referred to as the Finnish ranking) was launched in 2012 and currently has three levels for book publishers. From 2012 to 2014, book publishers were only distributed between two levels in a particular proportion. In essence, 90 per cent of book publishers were in level 1 and 10 per cent designated as level 2 (prestigious leading publishers). Level 3 was reserved for 25 per cent of level 2 journals and series (not book publishers). Since 2015, some book publishers have been awarded level 3.

It worth noting that book publishers and the book series they produce are ranked separately— which creates some confusion in the rankings; different countries deal with such issues differently. For example, the Norwegian model allows differences in levels such as a book series can be ranked as level 2, and its publisher ranked as level 1. On the contrary, the Finnish ranking determines that (even if a book publisher was ranked at level 1 before) it must be ranked at level 2 if one of its book series is assigned level 2. Despite this, the designers of the Finnish ranking warn: 'the quality levels applied in the Publication Forum predict the average quality and impact of large publication volumes, but they are too arbitrary a tool for the evaluation of individual publications or researchers' (Auranen & Pölönen 2012; Mañana Rodriguez & Pölönen 2018).

---

[16] The National Board of Scholarly Publishing  https://npi.nsd.no/organisering/npu?id=1109  accessed 16 April 2020
[17] The Bibliometric Research Indicator (BFI) provides an overview of research publications from Danish universities. https://ufm.dk/en/research-and-innovation/statistics-and-analyses/bibliometric-research-indicator accessed 16 April 2020
[18] *Publication Forum* is a classification of publication channels created by the Finnish scientific community to support the quality assessment of academic research.  https://www.julkaisufoorumi.fi/en

## 3.4. The List of Scientific Publishers in Poland

In Poland, the book evaluation system was quantitative, having some formal mandatory criteria (e.g. the length of a monograph); however, neither the prestige of book publishers nor peer-review, book reviews, citations, or book visibility was measured. According to Kulczycki (2018), 'a key weakness of the Polish system' was that it gives the same number of points for books published by Cambridge University Press, books published by a small local publisher, and even self-published books.

Inspired by foreign publisher rankings, the List of Scientific Publishers was announced in Poland in 2018 (Kulczycki 2019). During the initial phase, to identify the most prestigious (level 2) and the less prestigious (level 1) publishers, the Ministry of Science and Higher Education established "the Ministerial group for the lists of scientific journals and academic publishers". This advisory body, consisting of almost 20 academics, prepared the first publisher list based on several data sources: (1) the Finnish Publication Forum; (2) the Norwegian Register for Scientific Journals, Series and Publishers; (3) the Scholarly Publishers Indicators (Spain); (4) the Book Citation Index (Clarivate Analytics); (5) Scopus (Elsevier); (6) data from the Polish research evaluation system to identify publishers of books authored by Polish researchers in 2013-16.

In 2019, the Ministry of Science and Higher Education founded the Research Evaluation Council, consisting of over 30 academics responsible for research evaluation of Polish higher education and research institutions. Polish and foreign publishers interested in being included in the List of Scientific Publishers were invited to apply to the committee responsible for maintaining the list. The chair of this committee, Emanuel Kulczycki, informed us that the ministry could modify the list prepared by the academics if needed.

The latest Polish List of Scientific Publishers comprises 36 publishers at the highest level 2 and 677 publishers at the lower level 1[19]. For every book issued by a level 2 publisher, institutions earn 200 points in the sciences and 300 points in the social sciences and humanities. For every book produced by a level 1 publisher, institutions acquire 80 points in the sciences and 100 points in the social sciences and humanities. For every book published by any other publisher (not included in the List of Scientific Publishers), institutions obtain only 20 points. More information about the relation between points and funding for institutions is provided by Kulczycki (2017).

At first sight, Lithuania (a close neighbour of Poland) may seem to have a similar system which also distributes points according to a publisher's level. However, Lithuania introduced the List of Globally Recognised Publishers in 2006, soon after dropped it, and revived slightly different lists of prestigious publishers in 2017.

## 3.5. Prestigious publishers in Lithuania

Lithuania, being a small post-Soviet country, introduced a metric-based funding system in 2005, the year in which Norway launched its Norwegian Register. Since 2005, the Lithuanian regulations have defined 'prestigious publishers' as publishers which (1) continually release publications authored by national and international researchers, (2) distribute their products in many countries, (3) issue globally recognised journals and series of books, and (4) provide enough

---

[19] The order of Ministry of Science and Higher Education (Poland), 17 December 2020, https://www.bip.nauka.gov.pl/g2/oryginal/2019_12/9403b58df69c1fb6eee08a2225faf847.pdf accessed 17 November 2020

information about these achievements on their websites. According to the formal Lithuanian definition, prestigious publishers should fulfil all four listed above criteria.

In 2006, designers of Lithuania's national performance-based funding system changed some rules. They explained the aims of these amendments, declaring that they seek to incentivise research institutions to work efficiently, to raise their international competitiveness, and to comply with the needs of the state.[20] Within this order for outputs in the sciences (and not for the social sciences and humanities), policymakers introduced the List of Globally Recognised Publishers. It comprised sixteen named book producers such as Elsevier Science Group, Springer Group, Oxford and Cambridge University Presses, and other similar publishers.

We want to explicitly draw attention to the fact that publishers of book outputs in the sciences were assessed *based on their journal publishing activities*. The List of Globally Recognised Publishers concluded by stating that prestigious publishers of monographs are those publishers which issue at least *five peer reviewed journals* indexed in the Journal Citation Reports (Clarivate Analytics, the former Institute of Scientific Information).[21] Officially, the list was revoked soon after, in 2010.

Yet the List of Globally Recognised Publishers is still alive in Lithuanian researchers' minds. From 2004 to 2016, the experts (who are senior researchers in their fields) consistently scored monographs or edited volumes issued by those sixteen large publishers as the highest level book outputs produced by prestigious publishers.

From its beginning, the formal Lithuanian metric-based system was a purely quantitative formal assessment. However, in 2009, significant changes happened in the Lithuanian metric-based funding system. The evaluation was divided into two parts: *peer review assessment* (for 20 per cent of institutional outputs) and *formal metric-based assessment* (for the remaining eligible pieces of research). The Research Council of Lithuania administers this ex-post evaluation. It assigns self-registered senior researchers (henceforth referred to as the experts) into the pool for two expert panels (one for the sciences and the other for the social sciences and humanities). The experts have not been named; they work anonymously. According to the Lithuanian regulations, in the peer review assessment part, the experts evaluate the quality of the research presented in a book; in the formal metric-based assessment part, the same experts should appraise the prestige of book publishers.

## 4. Inconsistencies in assessing the prestige of book publishers

In Lithuania, for research outputs in the sciences, only monographs and chapters in edited volumes produced by prestigious publishers are eligible for formal assessment and for being assigned to the only existing level, prestigious (otherwise, they are rejected and not scored). In contrast, for the outputs in the social sciences and humanities, scholarly books issued by any publishers are eligible, but books produced by prestigious publishers (level 2) earn several times more points than those at level 1.

---

[20] The Ministry of Education and Science of the Republic of Lithuania, 12 April 2006 order No ISAK-685 (Valid from 7 May 2006, not valid from 19 July 2009) https://www.e-tar.lt/portal/lt/legalAct/TAR.94C0DCC94B9A accessed 17 November 2020

[21] The Ministry of Education and Science of the Republic of Lithuania, 29 June 2009 order No ISAK-1321 (Valid from 19 July 2009, not valid from 21 July 2010) https://www.e-tar.lt/portal/lt/legalAct/TAR.3BF7B2B40A59

When setting out to investigate the Lithuanian case, we selected publishers for which experts did not agree on their prestige. In this way, we identified sixteen inconsistently assessed publishers (and thirteen self-published cases[22]) that issued multiple books (either in the sciences or in the social sciences and humanities). This means the same publisher was sometimes assessed by anonymous experts as prestigious, while in other cases it was assessed as non-prestigious. These publishers issued a total of 224 books submitted for evaluation by the Lithuanian Research Council expert panels. The experts designated 93 books as published by prestigious publishers. They also rejected 63 books published by the same publishers, thus allocating these publishers both to the highest and to the lowest categories. The remaining 68 books produced by the same publishers were submitted to the not-so-prestigious publisher category (allowed only for books in the social sciences and humanities).

Table 1 shows the publishers ranked by experts as both prestigious and at least one other category, that is, not-so-prestigious (but sufficient for the social sciences and humanities) or not prestigious at all (usually for outputs in the sciences). Some publishers have consistently been classified as prestigious. We did not investigate these publishers in this study.

Table 1. Publishers awarded by the Lithuanian experts to the highest category and at least one other category in the national research funding assessments from 2004 to 2016, also the levels and years the same publishers were awarded in the Norwegian Register.

| The titles on publishers identified by ISBNs in the Global Register of Publishers (country ISBNs were registered) | The number of books evaluated by the experts and categories publishers received within the years | | | The scientific level of the publisher and the years from the Norwegian Register |
|---|---|---|---|---|
| | Prestigious publisher (for all sciences) ** | Not-so-prestigious, lower category (for SSH, scored) | Not prestigious, outcomes rejected (for the sciences) | |
| In Tech d.o.o. (Croatia) | 16 \| 2010–2011 | 4 \| 2010–2016 | 25 \| 2012–2016 | level 1 \| 2007– |
| **OmniScriptum GmbH & Co. KG[4] (Germany)** | **20** | **26** | **18** | – |
| VDM Verlag Dr Müller* | 5 \| 2008–2011 | 3 \| 2007–2012 | – | level 1 \| 2008–2010 |
| LAP Lambert Academic Publishing* | 14 \| 2009–2011 | 17 \| 2012–2016 | 16 \| 2010–2016 | level 0 |
| Südwestdeutscher Verlag* | 1 \| 2010 | – | - | – |
| Scholars' Press* | – | 3 \| 2015 | 2 \| 2014 | level 0 |
| Palmarium Academic Publishing* | – | 1 \| 2014 | – | – |
| GlobeEdit* | – | 1 \| 2015 | – | – |
| Nova Science Publishers, Incorporated (the USA) | 9 \| 2009–2015 | 12 \| 2006–2016 | 9 \| 2012–2016 | level 1 \| 2005–2017 |
| Peter Lang (the USA) | 10 \| 2008–2016 | 10 \| 2007–2016 | 3 \| 2012 | level 1 \| 2004– |
| Authors or miscellaneous publishers (Austria, Australia, Estonia, Finland, Germany, Latvia, and Lithuania) | 2 \| 2009, 2011 | 9 \| 2004–2016 | 2 \| 2008, 2015 | – |
| Studium Press, LLC (the USA) | 1 \| 2014 | – | 2 \| 2013, 2014 | level 0 |
| Begell House Publishers, Incorporated (the USA) | 4 \| 2005–2010 | – | 1 \|2012 | level 1 \| 2004– |
| Herder-Institut e.V. (Germany) | 1 \| 2015 | – | 1 \|2005 | level 0 |
| IGI Global (the USA) | 8 \| 2011–2016 | 2 \| 2010, 2014 | 1 \| 2013 | level 1 \| 2007– |
| Shaker Verlag (Germany) | 1 \| 2017*** | – | 1 \| 2016 | level 1 \| 2005– |
| Cambridge Scholars Publishing (the UK) | 20 \| 2010–2016 | 4 \| 2010–2016 | – | level 1 \| 2006-2018 |
| Hermann (France) | 1 \| 2017*** | 1 \| 2013 | – | level 1 \| 2009– |

\* Publishers as they are named in the bibliographic data of books submitted to the assessment in Lithuania.
\*\* The Lithuanian level 'prestigious publishers' is equal to level 2 in the Norwegian Register.
\*\*\* The status of the publishers was taken from the orders of the Research Council of Lithuania.

---

[22] Using the GRP, thirteen self-publishers were identified based on their ISBNs. According to the Lithuanian formal regulations, self-published works do not qualify for submission to metric-based assessments, neither do those having no ISBNs or misleading ISBNs.

By checking the GRP, data on registrants of ISBNs reveals quite a striking finding. It appears that the publisher listed in a library catalogue (which shows the bibliographical information of a book which librarians usually take from the book's copyright page) and the registrant of the ISBN do not necessarily match. Apparent discrepancies between these are detailed below.

The first finding reveals that some publications (whose copyright pages state that the publishers are universities or academic institutions) were actually identified by the GRP as self-published. According to the GRP, the registrants of those books are author-publishers (or miscellaneous publishers), who may have been assigned just one or two ISBNs each. Of these thirteen self-published books, seven were positively assessed, with two ranked as published by prestigious publishers.

A second noteworthy finding was that a German-based company, OmniScriptum GmbH & Co KG, was indicated in the data as six separate publishers (compiled in the first column and marked with an asterisk in Table 1). Nevertheless, these imprints do not exist in the GRP, despite OmniScriptum having identified every brand separately on its webpage. The Lithuanian experts awarded two imprints—Verlag Dr Müller (VDM) and Lambert Academic Publishing (LAP)—the highest category from 2009 to 2011. Currently, both brands are included in the Beall's List of vanity presses[23].

Thirdly, book outputs produced by LAP were both rejected as inappropriate and awarded the highest category, within the same years. An in-depth investigation of the disagreements among the Lithuanian experts revealed a quite complicated situation. In 2009, both panels (in the sciences and the social sciences and humanities) decided that LAP was a prestigious publisher and book outputs received maximum points. Nevertheless, after a year, in 2010, the panel in the sciences rejected the book outputs issued by LAP as produced by a non-prestigious publisher (and institutions received zero points). At the same time, the panellists in the social sciences and humanities awarded the books issued by LAP the highest category (and institutions received the maximum points). Then, in the assessment of 2011, both panels decided that LAP was a prestigious publisher, and six books submitted in the sciences as well as in the social sciences and humanities received the maximum points again. A turning point occurred in 2012 when the submission of monographs published by LAP doubled. Regrettably for the institutions (which were aware that the experts treated LAP as a prestigious publisher in the preceding year), the experts on both panels decided that LAP was not prestigious anymore. So, book outputs in the sciences were rejected (meaning no points), while those in the social sciences and humanities were awarded level 1 (meaning fewer points).

There are more inconsistencies in the experts' decisions. The next example deals with the status of InTech d.o.o. The Lithuanian experts scored maximum points for sixteen outputs (in the sciences) published by InTech d.o.o. over 2010-11. However, from 2012 to 2016, the experts decided that 25 outputs produced by this publisher were inappropriate because a non-prestigious publisher issued them. Thus, these 25 were rejected, and institutions did not receive the points (and funding) they expected, even though they had been incentivised to publish with this book producer one year earlier. Nevertheless, Lithuanian institutions received their points for chapters (in four edited volumes published by the same InTech d.o.o.) as outputs in the social sciences and humanities at the lower category (not-so-prestigious publishers)—where any publisher is eligible.

---

[23] The Beall's List of vanity presses. *What is vanity press*? https://beallslist.net/vanity-press/

Significant changes in circumstances regarding a publisher's prestige have surrounded the widely known UK-based publisher Cambridge Scholars Publishing (formerly Cambridge Scholars Press Ltd.[24]). It is interesting to note that the Lithuanian experts designated twenty of its titles as being produced by a prestigious publisher from 2010 to 2016. At the same time, four books were classified as issued by a not-so-prestigious one (so institutions received fewer points). In 2018, the Lithuanian experts awarded Cambridge Scholars Publishing the highest level, deeming it a prestigious publisher in the humanities.[25]

There is significant controversy around Cambridge Scholars Publishing. In France, when describing the inevitable confusion about some publishing houses' value, interviewed researchers mentioned Cambridge Scholars Publishing:

> 'I published there, so I found it quite good, but lately I learnt from English researchers that they consider it their Harmattan [...] Harmattan is not greatly considered by "serious" French researchers' (Williams & Galleron 2016).

However, in the UK, Cambridge Scholars Publishing was among 39 publishers which had twenty or more books submitted to the Research Excellence Framework across the arts and humanities in 2014 (Tanner 2016). In the UK REF2014, nearly three hundred outputs (authored book, edited books, chapters in books) produced by this publisher were selected as institutional choices of their excellence.[26]

In Norway, Cambridge Scholars Publishing had level 1 status, and it was ninth on the list of top ten publishers, covering 25 per cent of all scholarly book outputs published in international languages in the social sciences and humanities between 2005 and 2009 (Sivertsen & Larsen 2012). However, Cambridge Scholars Publishing received level 0 status in 2019, although Norwegian scholars still publish their works with this publisher, which is confirmed by significant numbers of production points registered in the Norwegian Publication Indicator[27].

In the Danish BFI lists, Cambridge Scholars Publishing first appeared in 2011 (BFI lists had no levels for book publishers at that time), then it became a level 1 publisher over 2012-13, and has disappeared from the BFI lists since 2014. In Finland, researchers on panels also assigned Cambridge Scholars Publishing to the basic level 1 status. In the Flemish database (VABB-SHW), this book producer is indicated as a level 0 publisher, having only some ISBN titles peer reviewed.

Another striking instance occurred in the Lithuanian data when a publisher became prestigious in the sciences category within a year. In 2017, the unnamed experts on the sciences panel rejected a monograph published (in 2016) by Germany-based publisher Shaker Verlag on the basis that the publisher was not prestigious. In 2018, the experts (we do not know if these were the same anonymous experts) selected Shaker Verlag as prestigious in the sciences.[28] In Norway, Finland,

---

[24] Cambridge Scholars Publishing Ltd in the Companies in the UK. Copyright © Comdevelopment Ltd 2020 https://www.companiesintheuk.co.uk/ltd/cambridge-scholars-publishing

[25] The Research Council of Lithuania 31 October 2019 order No V-554, the list of prestigious publishers in the humanities https://www.lmt.lt/lt/doclib/w9ytfmttnaqy9ucz2mkwwnaekvghfgym

[26] REF2014, Research outputs (REF2) https://webarchive.nationalarchives.gov.uk/20170302140351/http://results.ref.ac.uk/DownloadSubmissions/ByForm/REF2 accessed 20 April 2020

[27] Cambridge Scholars Publishing in The Norwegian Publication Indicator https://npi.nsd.no/forlagoversikt/forlaginfo/faglig?forlagId=19631

[28] The Research Council of Lithuania 31 October 2019 order No V-556, the list of prestigious publishers in the sciences https://www.lmt.lt/lt/doclib/e9wcjewhaaukpu981fwqz14k6kg2jfq8 accessed 20 April 2020

and Denmark, this book producer has a level 1 publisher status. In the Flemish database, Shaker has a level 0 publisher status.

Since 2018, the Research Council of Lithuania has distributed three separate lists of book publishers determined as prestigious in the sciences (nine publishers), the social sciences (eleven publishers)[29], and the humanities (twenty-three publishers) on its website. However, this does not explain if books produced by these prestigious publishers in subsequent years would receive maximum points as well—uncertainties for the submitting institutions still exist.

Additionally, the results presented in the last column of Table 1 indicate if the publisher was ranked in the Norwegian Register and if so, the years it happened. Level 0 means that somebody registered the publisher (or the publisher did so). However, they were not assigned a level (the Norwegian Centre for Research Data examines whether the book channel meets the minimum requirements). The sign '–' indicates that nobody has submitted the publisher to the Norwegian Register.

Notably, not a single publisher ranked as prestigious in Lithuania appeared at the Norwegian Register's level 2, which is Norway's highest level, and four publishers ranked as prestigious in Lithuania did not attain the minimum (level 1).

In sum, several cases vividly illustrate that the prestige of publishers relies on the impressions or previous experience of the scholars who serve on assessment panels. A considerable disagreement has persisted among researchers and experts as to the prestige of publishers, not only in different countries but even within the same country (i.e. Lithuania). The evidence presented above challenges what may be defined as a prestigious publisher.

The following section is an analysis of whether the national Norwegian and Lithuanian regulations are clear enough to determine the required status of publishers (e.g. academic or prestigious publisher, respectively). If so, scholars could easily follow the rules, and policymakers would reach their policy goals—to create an incentive for academics to publish their research with the most prestigious publishers.

## 5. Verifying whether book publishers meet necessary prerequisites

We reviewed the above-analysed international publishers listed in Table 1 to see whether we could verify that publishers meet *the minimum requirements for entry* into the Norwegian Register and the Lithuanian *formal definition of prestigious publishers* set in the regulations.

The importance of making the right decision regarding book publishers is more prominent for scholars in countries approaching metric-based assessment systems (e.g. Norway or Lithuania) than in countries having peer review evaluation (e.g. the UK). As an example, if Lithuanian researchers published with the 'wrong' publishers, institutions would not earn funding for research (currently this system is used to allocate almost half of the governmental funds for research).

Table 2 shows data on publishers anyone could collect from various sources. The first column shows how transparent and identifiable publishers are (to be discussed in Sub-section 5.1). The middle columns give the minimum necessary prerequisites for inclusion into the Norwegian,

---

[29] The Research Council of Lithuania 31 October 2019 order No V-555, the list of prestigious publishers in the social sciences https://www.lmt.lt/lt/doclib/grc78rw5k5pk3tyskcpm8bguunhwt323

Finnish, and Danish rankings of publishing channels (to be discussed in Sub-section 5.2). The last section offers some insight into how the publishers fit the Lithuanian explanation of 'prestige' (to be discussed in Sub-section 5.3). To be exact, we assumed that if publishers are international, they must provide information about the distribution of their books, their policies, and their authors on their websites in internationally understandable languages.

Table 2. Publishers which Lithuanian panel experts awarded the highest and at least one other category compared with their ranks in the metric-based assessments by various other countries
(data as of 19 April 2019)

| The transparency of the publishers | The minimum requirements for entry to the national registries, and the levels publishers awarded by 2019 | | | | | | | The Lithuanian description of prestigious publishers | |
|---|---|---|---|---|---|---|---|---|---|
| 'Registrant name' in the Global Register of Publishers \| 'Publisher' on the website (if different) \| a country of ISBNs, declared, year established | Policy on peer review of books | Editorial/ Advisory Board | Author- ship | Publisher's level in the national system in | | | | Publish book series or journals | Languages of the content provided on the website |
| | | | | Norway | Finland | Denmark | Flanders | | |
| 1 | 2 | 3 | 4 | 5 | 6 | 7 | 8 | 9 | 10 |
| In Tech d.o.o. (Croatia, 2007) = IntechOpen (the UK, 2017) | **yes** \| COPE* | yes | Int'l | level 1 2007– | level 0 | – | – | book series | English |
| OmniScriptum GmbH & Co. KG (18 academic brands, Germany, 2002) | | | | | | | | | |
| VDM Verlag | – | – | Int'l | level 1 2008-10** | level 1 2012-14 | – | – | ? | English |
| LAP Lambert Academic Publishing | – | – | Int'l | level 0 | level 0 | – | – | ? | English |
| Nova Science Publishers**** (the USA, 1985) | **yes** | ? ***** | Int'l | level 1 2005-17 | level 1 2012-19 | level 1*** 2008-19 | ISBN-selection | journals, book series | English |
| Peter Lang (the USA, Switzerland, 1970) | **yes** | – | Int'l | level 1 2004-19 | level 1 2012-19 | level 1 2008-19 | ISBN-selection | journals, book series | English |
| Studium Press (the USA, India, 1980) | – | = | ? | level 0 | – | – | – | book series | English |
| Begell House (the USA, 1991) | – | ? | Int'l | level 1 2004-19 | level 1 2012– | level 1 2008-19 | – | journals, book series | English, six more languages |
| Herder-Institut (Germany, 1990) | – | – | ? | level 0 | level 1 2014-19 | – | – | journal, book series | German, English |
| IGI Global**** (the USA, 1998) | **yes** COPE | ? | Int'l | level 1 2007-19 | level 1 2012-15 and 2018-19 | level 1 2008-19 | to employ peer review for all books | journal, book series | English, Chinese |
| Shaker Verlag (Germany, 1986) | – | – | Int'l | level 1 2005-19 | level 1 2012-19 | level 1 2008-19 | ISBN-selection | book series | German, English, Dutch |
| Cambridge Scholars Publishing (the UK, 2001) | – | ? | Int'l | level 1 2006-18 | level 1 2012-19 | level 1 2011-13 | ISBN-selection | book series | English |
| Hermann (France, 1876) | – | – | ? | level 1 2009-19 | level 1 2015-19 | level 1 2008-19 | – | book series | French |

\* Publisher is a member of the Committee on Publication Ethics (COPE).
\*\* as VDM Verlag Dr Müller Aktiengesellschaft & Co. KG. in the Norwegian Register. VDM Verlag relaunched as OmniScriptum in 2013.
\*\*\* The BFI list of publishers in Denmark had no levels for book publishers from 2008 to 2010. Two levels (1 and 2) for book publishers have launched since 2012.
\*\*\*\* Publishers included into the Beall's List of vanity presses https://beallslist.net/vanity-press/ assessed 9 June 2020.
\*\*\*\*\* '?' means that we were not able to find information on authorship or about an editorial/advisory board.

## 5.1. Misdemeanours in the appearance of publishers

To interrogate the data further, we chose the Principles of Transparency and Best Practice in Scholarly Publishing developed by well-known scholarly organisations[30] as the primary standard. As an example, in a study on scholarly journals' compliance with this standard (Choi et al. 2019), the sixteen principles of the standard were sub-divided into 33 items in four different categories: (1) basic journal information, (2) publication ethics information, (3) copyright and archiving information, (4) profit model information. For the analysis presented in this sub-section, we adjusted the items proposed in the category 'basic journal information'.

---

[30] Principles of Transparency and Best Practice in Scholarly Publishing. Developed by the Committee on Publication Ethics (COPE), the Directory of Open Access Journals (DOAJ), the Open Access Scholarly Publishers Association (OASPA), and the World Association of Medical Editors (WAME). The third version published on 15 January 2018 http://wame.org/principles-of-transparency-and-best-practice-in-scholarly-publishing accessed 16 May 2020.

*ISBNs and publishers*. The regulations mandate ISBNs for book outputs in Norway, Finland, Denmark, Flanders, and Lithuania. According to the International ISBN Agency, it is always the publisher of the book who should apply for the ISBN[31]. Thus, we examined the transparency of publishers, comparing the 'registrant name' in the GRP with the 'publisher' as it appears on the website it owns. We supposed that the publishers are transparent for authors, readers, and evaluators when this information matches.

The data show that two (out of eleven) publishers have a different presence on their webpages and the GRP: In Tech d.o.o. (named IntechOpen on its website) and OmniScriptum (which has numerous brands recognised on its website).

The publisher In Tech d.o.o. would look inconsistent for some researchers and the panel experts who assess its book outputs because searches in the GRP (by the prefix '978-953-307' of Lithuanian outputs' ISBNs) produce the publisher In Tech d.o.o. based in Croatia. However, the URL (provided on the GRP) directs users to the IntechOpen website [32], which declares only its headquarters in the UK ('About IntechOpen'). Nevertheless, the 'Contacts' page reveals that IntechOpen has two offices: In Tech d.o.o. in Croatia (registered in 2007[33]) and IntechOpen Limited in the UK (registered in 2017[34]). Despite its achievements and membership of COPE and OASPA (both being developers of the principles of transparency), IntechOpen stands as a level 1 publisher in Norway, a level 0 in Finland, and has no level in Denmark and Flanders.

Researchers and assessment panel experts would have more doubts regarding the publisher OmniScriptum. In the Lithuanian dataset, OmniScriptum consists of six imprints. However, only two of them (those awarded the prestigious category in Lithuania and listed in the Norwegian Register) are included in Table 2. At its origins, it was VDM Verlag launched in 2002 and relaunched as 'OmniScriptum' in 2013. Lambert Academic Publishing (LAP) is another brand of OmniScriptum, about which searches on the internet reveal claims it is a predator, vanity press[35], or at least questionable (Broz & Stöckelová 2018).

Currently, OmniScriptum openly declares its policies and business models on its website:

> 'Yes, we are aware of the criticism towards OmniScriptum that can be found on the web. […] Our company has changed tremendously in the last years. We have changed our business (no more Wikipedia since ages), we have changed our publishing terms, we have even changed our name. Just to clarify – we are OmniScriptum! […] Meanwhile our publishing group incorporates more than 45 imprints'. [36]

---

[31] Who should apply for ISBN? International ISBN Agency https://www.isbn-international.org/content/what-isbn accessed 16 May 2020.
[32] About IntechOpen https://www.intechopen.com/about-intechopen accessed on 1 May 2020
[33] Fininfo https://www.fininfo.hr/Poduzece/Pregled/in-tech/Detaljno/107379 accessed on 1 May 2020
[34] Companies in the United Kingdom © Comdevelopment Ltd 2020 https://www.companiesintheuk.co.uk/ltd/intechopen accessed on 1 May 2020
[35] The Beall's List of vanity presses. *What is vanity press*? https://beallslist.net/vanity-press/
[36] Omniscriptum - diversity and innovation https://www.omniscriptum.com/ accessed on 17 November 2020

## 5.2. Compliance with necessary prerequisites

The criteria for inclusion of new scientific publication channels into the Norwegian Register[37] are like those required in Finland and Denmark. Book publishers should have (1) established procedures for external peer review and (2) an academic editorial board (or an equivalent) primarily consisting of academics; also, they should (3) issue books authored by an international or at least a national research community.

The first prerequisite—*necessary procedures for external peer review in book publishing*—is essential, as it usually takes place in journal publishing. However, independent academic book publishers operate differently (Derricourt 2012); this is why we looked for the policies on peer review practices on the publishers' websites (Table 2, column 2). We found that only four publishers make publicly available their statements or descriptions about their peer review procedures, the main requirement for publishers accepted for the entry level into these four publisher assessment systems.

The second prerequisite—*a required advisory board of academics*—is declared as a list of people only on the IntechOpen website (Table 2, column 3). The symbol '–' means that we did not find any advisory board on the publisher's website. The sign '?' means that publishers do not publish who is on their advisory board; instead, they list authors, editors, and reviewers (in some cases) in one general list (e.g. Nova Science Publishers, Begell House). Alternatively, Cambridge Scholars Publishing lists 130 boards in the physical sciences and 102 boards in the social sciences on its website.  It is challenging to conduct the assessment of the presence of an editorial or advisory board or scientific committee on the websites of book publishers because of the numerous practices book publishers have in place.

The third prerequisite—*an international or at least a national authorship*—is required from book publishers for entry and standing at the minimum level 1 (Table 2, column 4). According to the Norwegian Register requirement, international publishers should have less than two thirds of their authors from the same country, while national authorship means 'no single institution is responsible for more than two-thirds of the publications in the channel over time'.

Because we did not find a single piece of advice on the source of calculation in the NPI website, we just checked the publishers' websites to see their online book catalogues or statements on their authorship. For example, IntechOpen states:

> 'Our community ranges from key opinion leaders of the international academic and scientific community, including Nobel Laureates and the top 1% of the world's most cited authors, to the next up-and-coming generation of scientists looking to make their mark.'
> https://www.intechopen.com/about-intechopen

The symbol '?' means that there were no authorship statistics or relevant information on the publisher's website. For example, the Herder Institute has no online book catalogue, statement, or lists of authors or editors. However, we found that it is a unit of the Leibniz Association (the Leibniz-Gemeinschaft), having 96 non-university research institutes. Thus, we could suppose that the Herder Institute would have at least a national authorship and meet this entry-level requirement. Nevertheless, we are not sure if we could make such an assumption about the fulfilment of this mandated criterion. So, it is a challenging task for assessors of book publishers to identify the level of publishers' authorship.

---

[37] In the Norwegian Register, the procedures for processing new submissions include: "... New scientific publication channels can be submitted continuously. ... Submissions from commercial publishers will not be considered." https://dbh.nsd.uib.no/publiseringskanaler/OmProsedyrer accessed on 17 November 2020

Additionally, we collected data on *the levels the publishers gained in the national systems* to compare our findings on the fulfilment of compulsory requirements with the publishers' levels (Table 2, columns 5-8). There are some disparities in the ranks of IntechOpen and the Herder Institute in their rankings in Norway and Finland, and it is not clear if these publishers had actually been considered and had not received any level in Denmark. However, LAP and Studium Press were not approved unanimously in all countries. The results suggest that some publishers (e.g. Begell House, Cambridge Scholars Publishing, Hermann, or Shaker Verlag) have no verifiable mandatory prerequisites in place.

Unfortunately, in some cases, we could not find a straightforward way to verify if the book publisher complies with the minimum necessary prerequisites; thus, transparent verification of compliance with the requirements is not possible.

Notwithstanding, some publishers without exceptionally high results in other countries were designated as prestigious in Lithuania, which prompts examination of the formal national definition of prestigious publishers.

### 5.3. Adherence to the formal Lithuanian definition

The Lithuanian regulations of the formal assessment define *prestigious publishers* as publishers which continually (1) release publications authored by national and international researchers and (2) distribute their products in many countries. Moreover, prestigious publishers are classified as such when they (3) issue globally recognised journals and book series, and (4) provide sufficient information about these achievements on their websites.[38]

The first feature—*international authorship*—is a similar prerequisite for level 2 or level 3 publishers in the above-discussed national rankings (in Norway, Denmark, and Finland), which define this as less than two thirds of authors from the same country (Table 2, column 2). Nonetheless, Lithuanian policymakers do not specify that authorship should be 'international'.

The second attribute—*distribution of books in many countries*—does not make publishers unique because currently, many publishers (and all we investigated in this study) distribute books they produce through their own websites, Amazon, or other vendors. Hence, we did not include this piece of additional information in the table.

The third quality—*issue globally recognised journals and series of books*—is somewhat ambiguous. It seems strange to judge book publishers according to their journal activities and decide whether their journals and book series are globally recognised (because the formal regulation does not explain how to measure the level of recognition). Likewise, uncertainty is left regarding publishers which issue only book series (and no journals) as to whether such publishers can be prestigious or not. Therefore, the results compiled in column 9 (Table 2) suggest that the production of a book series would be enough for publishers to be awarded the prestigious rank.

Notably, LAP (an imprint of OmniScriptum) announces on its website that its main targets are theses and dissertations (and we did not recognise series on its website). Nonetheless, the experts scored books they produced as those published by prestigious publishers, and with such a decision, they have created incentives for researchers to publish with this publisher.

---

[38] The Ministry of Education and Science of the Republic of Lithuania, 4 October 2017 order No V-747 (Valid since 1 November 2017) https://www.e-tar.lt/portal/lt/legalAct/69270ef0a8d411e78a4c904b1afa0332

The concluding requirement—*provide sufficient information about all achievements on their website*s—seems rational because achievements (if the formal regulations specify them) would help identify prestigious publishers. However, many controversial or questionable publishers have such perfect-looking websites that even experienced scholars do not recognise their failings.

Therefore, we decided that if international authorship is a mandatory feature, it would be reasonable to check whether the publishers have their policies and other content on their websites in English. Thus, any potential author could become acquainted with information provided on the website before submitting a manuscript, or the experts could ascertain that the publisher fits the definition of a prestigious publisher. Column 9 in Table 2 shows the languages of the content provided on the publisher's website. Only Hermann, the oldest publisher on the list, would not meet the fourth criterion because its webpage is only in French, so it is difficult to learn more about its achievements (and Google translate did not help in this case). Nevertheless, the Lithuanian experts designated Hermann as a prestigious publisher.

Given all the circumstances mentioned above, it is not surprising that the Lithuanian panels of experts have no consensus on the 'prestige' of book publishers (presented in Table 1). Even the official national regulations do not help to differentiate the prestigious from the not-so-prestigious publishers.

## 6. Discussion and conclusions

The evaluation of book outputs is debated at length in research papers studying national research assessment systems or examining indicators for the assessment of books.

In countries performing national research assessment exercises that rely on qualitative peer review assessment, researchers debate the benefits of a metrics-based approach versus a peer review approach (Allen & Heath 2013) or look for new ways to assess books (Basso et al. 2017). Meanwhile, UK policymakers, exploring ways to extend the possibilities for evaluating books, have introduced some pioneering prerequisites such as open access for monographs, which have already been widely discussed (Crossick 2015; Lockett 2018).

At the same time, in countries that use a quantitative assessment system, researchers express concerns about the effects of metrics-based research assessment on research practice in general. For instance, some institutions reward individual researchers using metrics that were originally intended to be used only at the institutional level (Aagaard 2015; Hammarfelt & de Rijcke 2015). Mouritzen and Opstrup (2020) reveal components of the Danish Bibliometric Research Indicator leading to gaming the system, and Rowlands and Wright (2019) investigate the effects of research assessment on research practice in Denmark. There seems to be relatively little debate about the underlying causes and consequences of incentive structures in countries that employ publishers' rankings to assess books.

Our findings show that no matter whether countries employ qualitative peer review or quantitative metrics-based assessment, they still use experts' knowledge to evaluate book outputs. In a qualitative approach, experts assess the submitted books individually (and according to the formal policies, they do not judge publishers). In contrast, in a metrics-based evaluation, empowered researchers select the most prestigious book publishers from all publishers that are considered to meet certain basic entry requirements.

As our examination of various national assessment systems reveals, publishers' rankings rely on four main prerequisites—ISBN prefix, external peer review procedures, a scientific publishing programme or advisory board, and national or international authorship. However, as our findings show, rankings do not disclose the details of their approval procedures, and neither do book publishers make the relevant information publicly available. Hence, there is little transparency in the process of determining whether publishers meet the minimum requirements for entry into a national register.

Furthermore, our results show that experts in different countries may have contradictory opinions on the prestige of a publisher. The same publisher may be ranked differently in the Danish, Finnish, Norwegian, and Lithuanian registers. Also, within Lithuania, the same publisher may be categorised as prestigious in one year and as satisfactory or even ineligible in the next year. These findings indicate that it is difficult to reach a common understanding of what it means to be a prestigious publisher. This raises doubts about the outcomes of assessments of books based on a judgement about their publisher.

It strikes us that neither quantitative nor qualitative assessment approaches stress the importance of the dissemination of research published in books. Rankings of book publishers are focused on the gatekeeping and content quality expected from publishers. They do not assess how publishers contribute to disseminating academic research and scholarship (e.g. through digital formats and high-quality metadata). Even though the Lithuanian regulations mention dissemination as a feature of prestigious publishers, there is no further explanation of what is meant or required.

Future research may focus on developing improved approaches for assessing books. Our suggestion would be to start from the idea that there are more roles of book publishers which are very important for communication between researchers: (1) quality control (e.g. peer-review, copyediting); (2) production (e.g. print runs, digital format, editions); (3) dissemination and archiving (e.g. metadata, digital formats, and long term digital preservation); (4) marketing (e.g. book reviews, social media); and other roles (*Future of scholarly publishing and scholarly communication* 2019).

Publishers may decide how much they want to offer in each of the areas mentioned above. However, they need to be transparent about the services delivered in each stage for every book. Presumably, different publishers will make different choices in that respect. If publishers are transparent and indicate what they offer in terms of quality control, dissemination, and other services, national assessment systems can use this information. Ideally, the data is provided at the level of individual books so that there is no need to rely on general information about publishers.

While the evaluation of book outputs is debated at length in the literature, there is a lack of studies taking more detailed perspectives on editorial and peer review practices in book publishing, publishers' services and imprints, or book metadata. Future research may adopt a more in-depth attitude on practicalities in book publishing in which these factors are taken into account from research assessment perspectives. Another direction for future research is to develop a deeper understanding of why different countries take different approaches to evaluate book outputs. Understanding differences or similarities between countries may help explain why some countries take Open Access for books as a strategy for research assessment (Crossick 2015) while others developing a global and multilingual register[39].

## Acknowledgements

First, I am grateful for the exceptional support of my supervisor, Ludo Waltman. His knowledge and exacting attention to this research project have improved this study in innumerable ways.

The cleansing of the primary bibliographical data on Lithuanian book outputs, which consisted of many missing or incorrect ISBNs codes, would not have been successful without the generous assistance of Aldona Barodicaitė, head of the Lithuanian ISBN and ISMN Agency. The finishing touches were added by Stella Griffiths, executive director of the International ISBN Agency, who helped solve several misleading cases and explained the subtleties of the ISBN world. Stella and Aldona increased my confidence in analysing the final empirical data and prompted new research topics that I am currently working on.

I am also thankful to Julie Martyn, a professional language editor at Grammarfun. She kept this paper on track from the first draft prepared just to its final version.

---

[39] Academic Book Publishers (ABP): a global and multilingual register https://enressh.eu/wp-content/uploads/2019/08/Academic-Book-Publishers-ABP-A-global-and-interactive-register.pdf accessed 11 July 2021